\documentclass[twocolumn,english,pra,showpacs]{revtex4}
\usepackage[T1]{fontenc}
\usepackage[latin9]{inputenc}
\usepackage{amsmath}
\usepackage{amssymb}
\usepackage{graphicx}
\usepackage{esint}
\usepackage{amsfonts}
\usepackage{array}
\usepackage{multirow}
\usepackage{amstext}
\usepackage{CJK}
\usepackage{babel}
\usepackage{babel}
\usepackage{babel}
\usepackage{subfigure}

\begin{document}

\title{Recoil effects of a motional scatterer on single-photon scattering in one dimension}

\author{Qiong Li  }

\affiliation{State Key Laboratory of Theoretical Physics, Institute
of Theoretical Physics, Chinese Academy of Science, Beijing 100190,
China}

\author{D. Z. Xu }

\affiliation{State Key Laboratory of Theoretical Physics, Institute
of Theoretical Physics, Chinese Academy of Science, Beijing 100190,
China}

\author{C. Y. Cai }

\affiliation{State Key Laboratory of Theoretical Physics, Institute
of Theoretical Physics, Chinese Academy of Science, Beijing 100190,
China}

\author{C. P. Sun }

\affiliation{Beijing Computational Science Research Center, Beijing
100084, China}

\begin{abstract}
The scattering of a single photon with sufficiently high energy can
cause a recoil of a motional scatterer. We study its backaction on
the photon's coherent transport in one dimension by modeling the
motional scatterer as a two-level system, which is trapped in a
harmonic potential. While the reflection spectrum is of a single
peak in the Lamb-Dicke limit, multi-peaks due to phonon excitations
can be observed in the reflection spectrum as the trap becomes
looser or the mass of the two-level system becomes smaller.
\end{abstract}


\maketitle

\section{Introduction}

To realize all-optical devices at the single photon level for
quantum information processing, it is very crucial to control the
coherent transport of a single photon in some artificial quantum
architectures. Recent theoretical studies \cite{Fan,Zhou2008} have
demonstrated that, in one dimension, a fixed two-level system can
well serve this purpose in principle, since the interference between
the spontaneously emitted photon and the incident one can result in
a perfect reflection on resonance. Therefore, the TLS can act as a
quantum node to function as a quantum optical switch and a single
photon transistor
\cite{Lukin,Zhou2008-A,Zhou2009,Shi2009,Shi2011,Chang2011,Yan2011,Zhang2012,Huang2013}.
Here, we emphasize that this observation is made only for the
situation, in which the TLS is fixed and thus does not recoil in the
scattering process.

However, when the energy of the single photon becomes comparable to
the motional energy of the TLS, the recoil of the TLS caused by the
incident photon becomes evident. It is obvious that such a
backaction will have effect on the absorption and emission of
photons. For instance, the energy of the gamma-ray photon emitted by
a nucleus at rest is lower than the transition energy of the
nucleus, while that absorbed by a nucleus at rest is higher than the
transition energy of the nucleus, because in both emission and
absorption processes energy is lost into the recoil motion of the
nucleus. In an ensemble of free nuclei, the energy shift is so great
that the emission and absorption spectra have no overlap. As a
result, the nuclear resonance fluorescence cannot happen in free
nuclei.

In contrast, for nuclei bound in a solid crystal, there exist
recoil-free emission and absorption, and in consequence the nuclear
resonance fluorescence can be observed. This phenomena is called the
M$\ddot{o}$ssbauer effect \cite{Mossbauer}. Here, the inherent
mechanism to suppress the recoil lies in the fact that the
conservation of momentum is actually satisfied by the solid crystal
as a whole instead of by a single nucleus alone. In a recoil-free
scattering process based on the M$\ddot{o}$ssbauer mechanism, the
novel quantum optical phenomena, such as the collective Lamb shift
and the electromagnetically-induced-transparency (EIT)-like
phenomena \cite{Rohlsberger2010,Rohlsberger2012}, were displayed in
an ensemble of nuclei interacting with the intense coherent photons
from the synchrotron radiation light sources.

It is believed that there will emerge more exciting quantum optical
effects, as photons of much better coherence and of much greater
beam intensity shall be utilized, which are produced from a novel
generation of light sources, such as the hard X-ray or gamma-ray
free-electron laser (FEL). Stimulated by the great progress in the
high energy coherent light source
\cite{Nat2007,Nat2010,Nat2011,Rohringer,Flash,LCLS,SACLA,XFEL}, a
novel research field, the X-ray quantum optics \cite{X-ray} is
gaining momentum to be born in the near future. At that time, when
the high energy photons, with perfect coherence, interact with
nuclei (especially for free atoms in the in situ x-ray measurement
\cite{in-situ-2008,in-situ-2012}), the recoil effect will become
dominant to lead to new quantum phenomena beyond the
M$\ddot{o}$ssbauer effect. It is worthy to notice that the existing
theoretical studies \cite{Rohlsberger2004,DZXU} only work well for
the situation that the nuclei are perfectly fixed by the solid
crystal.

In this paper, we set up a fully quantum model to directly manifest
the main characteristics of the recoil effects in the single-photon
scattering by a motional scatterer. In our one-dimensional (1D)
model, the motional scatterer is modeled as a TLS confined by a
harmonic potential and assume the 1D photon field has a linear
dispersion relation, which is a good approximation of the usual 3D
case. The Lamb-Dicke parameter (LDP) \cite{LD}, which is
proportional to both the wave number of the resonant photon and the
spatial spread of the TLS's ground state wavefunction, quantifies
the coupling strength of the TLS's internal state to its motional
state, and thus controls the recoil strength of the TLS in the
photon scattering process. In the Lamb-Dicke limit when the LDP is
very small, the scattering process is recoil-free, and in the
reflection spectrum there is a single peak of complete reflection at
the resonance point. As the LDP becomes larger, the recoil effect in
the scattering process becomes more evident. A main characteristic
of the recoil effect is the phonon induced multi-peaks in the
reflection spectrum; moreover, the scattered photon splits into a
mixed state due to its entanglement with the TLS's motional state,
and each component of the mixed state has a separate frequency
shift.

This paper is organized as follows: In Sec. II , we describe the
model. In Sec. III, we obtain the scattering state. In Sec. IV, we
study the reflection (transmission) spectrum and analyze the recoil
effects. In Sec. V, main conclusions are summarized.

\section{Model Setup\label{sec:Model-Setup}}

\begin{figure}
\begin{center}

\includegraphics[bb=230 50 595 430, width=5cm]{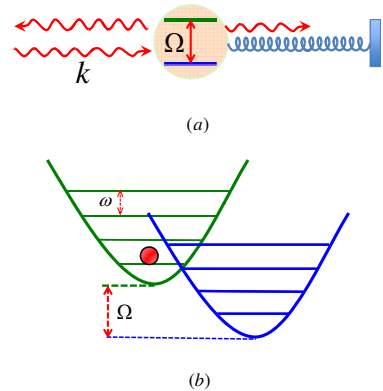}

\caption{\label{fig:setup} (Color online)  (a) Schematic of a single
photon with wave number $k$ incident on the motional two-level
system (TLS) with transition frequency $\Omega$. (b) The TLS is
trapped by a harmonic potential. Here, the upper green (lower blue)
harmonic oscillator represents the energy levels of the TLS in the
internal excited (ground) state. }
\end{center}

\end{figure}

\begin{figure}
\begin{center}
\includegraphics[bb=180 460 510 690, width=8cm]{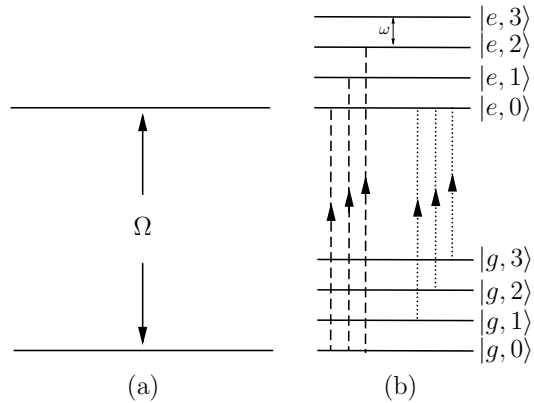}

\caption{\label{fig:level} (a) The energy level structure of the
fixed TLS. (b) The energy level configuration of the motional TLS,
including motional sidebands. }
\end{center}

\end{figure}

As schematically shown in Fig.\ref{fig:setup}, what we consider is a
1D system, consisting of a TLS, whose mass is $M$ and the internal
energy level spacing between the excited state $\left|e\right\rangle
$ and the ground state $\left|g\right\rangle $ is $\Omega$. The
spatial confinement of the TLS is modeled by a harmonic potential of
the oscillator frequency $\omega$, and therefore the motional state
of the TLS is denoted by a Fock state $\left|n\right\rangle $. Let
$a^{\dagger}$ ($a$) be the creation (annihilation) operator of the
vibrational quanta (phonons). Then the TLS's position (momentum)
operator reads as
\begin{eqnarray*}
x & = & \alpha(a^{\dagger}+a),\\
p & = & \frac{i}{2\alpha}(a^{\dagger}-a),
\end{eqnarray*}
with $\alpha=1/\sqrt{2M\omega}$. Therefore, the Hamiltonian
involving the motional TLS alone reads as
\begin{eqnarray}
\hat{H}_{a} & = & \Omega\left|e\right\rangle \left\langle
e\right|+(a^{\dagger}a+1/2)\omega,
\end{eqnarray}
and the corresponding energy level diagram is shown in
Fig.\eqref{fig:level}. While the fixed TLS has two levels, the
motional TLS has two sidebands made up of many sublevels. The state
of the motional TLS is denoted by $\left|g,n\right\rangle $ or
$\left|e,n\right\rangle $, which means the internal state is
$\left|g\right\rangle $ or $\left|e\right\rangle $ and the motional
state is $\left|n\right\rangle $.

The Hamiltonian of the 1D photon field is
\begin{eqnarray}
\hat{H}_{w} & = &
\int_{-\infty}^{+\infty}dk\omega_{k}c_{k}^{\dagger}c_{k},
\end{eqnarray}
with $\omega_{k}=\upsilon_{g}|k|$, where $k$ is the photon
wavevector and $\upsilon_{g}$ is the velocity of the confined photon
\cite{Fan2009}.

In the rotating wave approximation, the dipole coupling of the TLS
to the 1D photon field is described as
\begin{equation}
\hat{V}=J\int_{-\infty}^{+\infty}dk\left(c_{k}^{\dagger}\sigma_{-}e^{-i\alpha
k(a^{\dagger}+a)}+\text{h.c.}\right).\label{eq:V}
\end{equation}

Here, $J$ is the frequency-independent coupling between the photon
field and the TLS; $c_{k}^{\dagger}$($c_{k}$) is the creation
(annihilation) operator of the photon field, obeying the bosonic
commutation relations
$[c_{k},c_{k^{\prime}}^{\dagger}]=\delta_{k,k^{\prime}}$;
$\sigma_{+}$($\sigma_{-}$) is the internal transition operator, i.e.
$\sigma_{+}=\left|e\right\rangle \left\langle g\right|$,
$\sigma_{-}=\left|g\right\rangle \left\langle e\right|$;
$\exp\left[\pm i\alpha k(a^{\dagger}+a)\right]$ is known as the
displacement operator, which can creat a coherent state acting on
vacuum. In this model, the displacement operator causes transitions
between different motional Fock states. Thus, we name it the
transition operator here. The LDP, denoted as $\varepsilon_{LD}$, is
given by
\begin{equation}
\varepsilon_{LD}=\alpha k_{c},
\end{equation}
where $k_{c}$ is the resonant wave number, i.e.
$k_{c}=\Omega/\upsilon_{g}$.

\section{Single-Photon Scattering State }

A photon propagating in 1D will inevitably encounter the TLS. The
virtual absorption and emission of the photon by the TLS
dramatically affect the coherent transport of the incident photon,
which could be described as a single-photon scattering process by
the TLS.

A single photon, incident from the left side and with wave number
$k$, propagates in 1D, until it is scattered by the TLS, which is
initially in the lowest energy state. The input state for scattering
is  $\left|\phi_{k,0}\right\rangle
=c_{k}^{\dagger}\left|\text{vac}\right\rangle
\otimes\left|g,0\right\rangle $, where
$\left|\text{vac}\right\rangle $ represents the vacuum of the photon
field and $\left|g,0\right\rangle $ is the lowest energy state of
the TLS. In the scattering process, the TLS may virtually absorb and
spontaneously emit a photon, which transfers the single-excitation
from the photon field to the internal state of the TLS or vice
versa. Additionally, the vibrational motion of the TLS is disturbed,
and thus the number of phononic excitations may be changed.

In the subspace spanned by the basis $\left|g,m\right\rangle $ and
$\left|e,n\right\rangle $ ($m$ and $n$ are the phonon numbers), the
scattering state is assumed to be
\begin{eqnarray}
\left|\phi_{k,0}^{(+)}\right\rangle  & = & \sum_{n=0}^{\infty}\int_{-\infty}^{+\infty}dpu_{g}(p,n)c_{p}^{\dagger}\left|\text{vac}\right\rangle \otimes\left|g,n\right\rangle \nonumber \\
 &  & +\sum_{n=0}^{\infty}u_{e}(n)\left|\text{vac}\right\rangle \otimes\left|e,n\right\rangle .\label{eq:phik0-exact}
\end{eqnarray}
The wavefunctions $u_{g}(p,n)$ and $u_{e}(n)$ obey the Lippmann
Schwinger equation
\begin{eqnarray}
\left|\phi_{k,0}^{(+)}\right\rangle  & = &
\left|\phi_{k,0}\right\rangle
+\frac{1}{E_{k,0}-\hat{H}_{0}+i0^{+}}\hat{V}\left|\phi_{k,0}^{(+)}\right\rangle
,\label{eq:LS-exact}
\end{eqnarray}
where $\hat{H}_{0}=\hat{H}_{a}+\hat{H}_{w}$ and
$E_{k,0}=\omega_{k}+\omega/2$. It follows from
Eqs.(\ref{eq:phik0-exact}, \ref{eq:LS-exact}) that
\begin{eqnarray}
u_{g}(p,n) & = & \delta_{n,0}\delta(p-k)\nonumber \\
 &  & +J\sum_{m=0}^{\infty}u_{e}(m)\frac{\left\langle m\right|e^{-i\alpha p(a^{\dagger}+a)}\left|n\right\rangle }{E_{k,0}-\omega_{p,n}+i0^{+}},\label{eq:ugpn}\\
u_{e}(m) & = & \frac{J}{E_{k,0}-\omega_{e,m}+i0^{+}}\nonumber \\
 &  & \times\left[\left\langle m\right|e^{i\alpha k(a^{\dagger}+a)}\left|0\right\rangle +\sum_{n=0}^{\infty}F(m,n)u_{e}(n)\right],\nonumber \\
\label{eq:uem}
\end{eqnarray}
where $\omega_{p,n}=\omega_{p}+(n+1/2)\omega$,
$\omega_{e,m}=\Omega+(m+1/2)\omega$ and
\begin{eqnarray}
F(m,n) & = & \sum_{\bar{m}=0}^{\infty}\int_{-\infty}^{+\infty}dp\frac{J}{E_{k,0}-\omega_{p,\bar{m}}+i0^{+}}\nonumber \\
 &  & \times\left\langle m\right|e^{i\alpha p(a^{\dagger}+a)}\left|\bar{m}\right\rangle \left\langle \bar{m}\right|e^{-i\alpha p(a^{\dagger}+a)}\left|n\right\rangle .\nonumber \\
\label{eq:F}
\end{eqnarray}
Note that for motional TLS, the effective coupling to the photon
field is indeed the coupling constant $J$ times the vibrational
transition amplitude
\begin{eqnarray}
 &  & \left\langle m\right|e^{\pm i\alpha p(a^{\dagger}+a)}\left|n\right\rangle \nonumber \\
 & = & (\pm i)^{|m-n|}\sqrt{\frac{\left(\min\left[m,n\right]\right)!}{\left(\max\left[m,n\right]\right)!}}\nonumber \\
 &  & \times e^{-\alpha^{2}p^{2}/2}(\alpha p)^{|m-n|}L_{\min[m,n]}^{|m-n|}(\alpha^{2}p^{2}),\label{eq:transition}
\end{eqnarray}
where $L_{\min[m,n]}^{|m-n|}$ is the generalized Laguerre
polynomial.

\section{Single-Photon Scattering Probability}

Now we calculate the single-photon scattering probability, according
to the scattering state obtained in the above section. To this end,
we first reform the scattering state in the coordinate
representation as
\begin{eqnarray}
\left|\phi_{k,0}^{(+)}\right\rangle  & = & \sum_{n=0}^{\infty}\int_{-\infty}^{+\infty}dxu_{g}(x,n)c^{\dagger}(x)\left|\text{vac}\right\rangle \otimes\left|g,n\right\rangle \nonumber \\
 &  & +\sum_{n=0}^{\infty}u_{e}(n)\left|\text{vac}\right\rangle \otimes\left|e,n\right\rangle ,\label{eq:phik}
\end{eqnarray}
where
\begin{eqnarray}
c^{\dagger}(x) & = & \int_{-\infty}^{+\infty}\frac{dp}{\sqrt{2\pi}}c_{p}^{\dagger}e^{-ip\cdot x},\\
u_{g}(x,n) & = &
\int_{-\infty}^{+\infty}\frac{dp}{\sqrt{2\pi}}e^{ip\cdot
x}u_{g}(p,n).\label{eq:ugxnd}
\end{eqnarray}
From Eq.\eqref{eq:ugpn} and Eq.\eqref{eq:ugxnd}, we can obtain the
photon wavefunction $u_{g}(x,n)$. Actually, to study the scattering
problem, we only need to know the wavefunction in the limit
$|x|\rightarrow\infty$, where it is explicitly written as
\begin{eqnarray}
u_{g}(x,n) & \sim & \theta(-x)\left(\frac{e^{ik\cdot x}}{\sqrt{2\pi}}\delta_{n,0}+\frac{e^{-ik_{n}\cdot x}}{\sqrt{2\pi}}r_{n}(k)\right)\nonumber \\
 &  & +\theta(x)\frac{e^{ik_{n}\cdot x}}{\sqrt{2\pi}}t_{n}(k)\label{eq:ugxn-inf}
\end{eqnarray}

with $k_{n}=k-n\omega/\upsilon_{g}$ and
\begin{eqnarray}
t_{n}(k) & = & \delta_{n,0}+\theta(k_{n})\left(-i2\pi J/\upsilon_{g}\right)\nonumber \\
 &  & \times\sum_{m=0}^{\infty}u_{e}(m)\left\langle n\right|e^{-i\alpha k_{n}(a^{\dagger}+a)}\left|m\right\rangle ,\label{eq:tnk}\\
r_{n}(k) & = & \theta(k_{n})\left(-i2\pi J/\upsilon_{g}\right)\nonumber \\
 &  & \times\sum_{m=0}^{\infty}u_{e}(m)\left\langle n\right|e^{i\alpha k_{n}(a^{\dagger}+a)}\left|m\right\rangle .\label{eq:rnk}
\end{eqnarray}

From Eq.\eqref{eq:phik}, we observe that the photon is in a mixed
state, due to its entanglement with the TLS. Let $\phi_{n}(x)$
represent the photon wavefunction of the $n$th pure state component.
Obviously, there is $\phi_{n}(x)=u_{g}(x,n)$. For the $n$th pure
state component, the reflectance is $\left|r_{n}(k)\right|^{2}$ and
the transmittance is $\left|t_{n}(k)\right|^{2}$. The total
reflectance (transmittance) $R_{k}$ ($T_{k}$) is a statistical sum
of each pure state component, namely,

\begin{eqnarray}
R_{k} & = & \sum_{n=0}^{\infty}\left|r_{n}(k)\right|^{2},\label{eq:Rk}\\
T_{k} & = &
\sum_{n=0}^{\infty}\left|t_{n}(k)\right|^{2}.\label{eq:Tk}
\end{eqnarray}

According to
Eqs.(\ref{eq:uem},\ref{eq:tnk},\ref{eq:rnk},\ref{eq:Rk},\ref{eq:Tk}),
we numerically obtain the total reflectance (transmittance), from
which the probability current conservation, i.e.  $T_{k}+R_{k}=1$,
is confirmed.

\subsection{The Lamb-Dicke limit}

\begin{figure}
\begin{center}
  \includegraphics[bb=0 0 385 270, width=8cm]{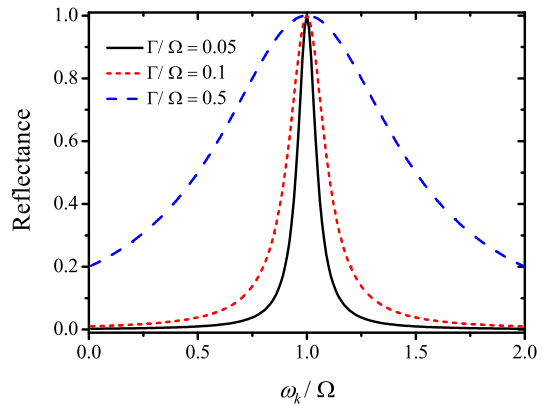}

\caption{\label{fig:LDlimit} (Color online) The reflection spectra
with $\Gamma/\Omega=0.05$, $0.1$ and $0.5$ are compared in the
Lamb-Dicke limit.}
\end{center}
\end{figure}

The recoil effects in the single-photon scattering arise from the
vibrational transition operator $\exp\left[\pm i\alpha
k(a^{\dagger}+a)\right]$ and the recoil strength depends on the LDP.
In the Lamb-Dicke limit, $\alpha k_{c}\rightarrow0$, the vibrational
transition operator can be expanded to the lowest order, i.e.
$\exp\left[\pm i\alpha k(a^{\dagger}+a)\right]\simeq1$.
Consequently, the recoil vanishes and the model reduces to the ideal
case \cite{Fan}. The result in the Lamb-Dicke limit can be obtained
by setting $\alpha=0$ in the above equations. In the limit
$\alpha\rightarrow0$, Eqs.(\ref{eq:uem}, \ref{eq:F}) become
\begin{eqnarray}
F(m,n) & = & \delta_{m,n}\theta(k_{n})\left(-\frac{i\Gamma}{J}\right),\\
u_{e}(m) & = &
\delta_{m,0}\frac{J}{\left(\omega_{k}-\Omega\right)+i\Gamma},
\end{eqnarray}
where the notation $\Gamma$ is defined as
\begin{equation}
\Gamma=2\pi J^{2}/\upsilon_{g}.\label{eq:Gamma}
\end{equation}
Further, the scattering wavefunction $u_{g}(x,n)$ is obtained as
\begin{eqnarray}
u_{g}(x,n) & = & \delta_{n,0}\left[\theta(-x)\frac{e^{ik\cdot x}}{\sqrt{2\pi}}+\theta(x)\frac{\omega_{k}-\Omega}{\omega_{k}-\Omega+i\Gamma}\frac{e^{ik\cdot x}}{\sqrt{2\pi}}\right.\nonumber \\
 &  & \left.+\theta(-x)\frac{-i\Gamma}{\omega_{k}-\Omega+i\Gamma}\frac{e^{-ik\cdot x}}{\sqrt{2\pi}}\right],\label{eq:ugxn_LD}
\end{eqnarray}
which shows that in the scattering process the motional state of the
TLS remains in the Fock state $\left|0\right\rangle $ and thus the
scattered photon is in a pure state. From Eq.\eqref{eq:ugxn_LD}, we
read the reflection (transmission) amplitude as
\begin{eqnarray}
t(k) & = & \frac{\left(\omega_{k}-\Omega\right)}{\left(\omega_{k}-\Omega\right)+i\Gamma},\\
r(k) & = & \frac{-i\Gamma}{\left(\omega_{k}-\Omega\right)+i\Gamma},
\end{eqnarray}
which further give the reflectance (transmittance) as
\begin{eqnarray}
T_{k} & = & |t(k)|^{2}=\frac{\left(\omega_{k}-\Omega\right)^{2}}{\left(\omega_{k}-\Omega\right)^{2}+\Gamma^{2}},\\
R_{k} & = &
|r(k)|^{2}=\frac{\Gamma^{2}}{\left(\omega_{k}-\Omega\right)^{2}+\Gamma^{2}}.
\end{eqnarray}
This result is in accord with Ref.\cite{Fan}. The probability
current conservation, i.e. $T_{k}+R_{k}=1$, holds. Therefore, we are
only concerned with the reflectance $R_{k}$. We plot the reflection
spectrum $R_{k}$ in Fig.\ref{fig:LDlimit}. It is seen that complete
reflection occurs at the resonant point $\omega_{k}/\Omega=1$, and
$\Gamma/\Omega$ is the width of the reflection peak. When the
coupling strength $J$ is very weak, a sharp reflection peak is
obtained.

\subsection{The recoil effects}

As stated above, the recoil strength is fixed by LDP. In the limit
$\alpha k_{c}\rightarrow0$, the recoil vanishes; as $\alpha k_{c}$
increases, the recoil becomes more important. In the Lamb-Dicke
regime $\alpha k_{c}\ll1$, let us expand the vibrational transition
operator in $\alpha$ as
\begin{eqnarray}
e^{-i\alpha k(a^{\dagger}+a)} & = & 1-i\alpha k(a^{\dagger}+a)\nonumber \\
 &  & +\frac{1}{2}(-i\alpha k)^{2}(a^{\dagger}+a)^{2}+O(\alpha^{3}).
\end{eqnarray}
If we keep only the first order, $\exp\left[\pm i\alpha
k(a^{\dagger}+a)\right]\simeq1$, the TLS may absorb the photon by
internal transition, but its vibrational motion is not affected.
Therefore, there is only one resonance at $\omega_{k}=\Omega$. If we
keep up to the second order, $\exp\left[\pm i\alpha
k(a^{\dagger}+a)\right]\simeq1\pm i\alpha k(a^{\dagger}+a)$, the TLS
may absorb the incident photon and meanwhile, its vibrational motion
gains a phonon or remains in the lowest sublevel. As a result, the
outgoing photon is in a mixed state, of which a component has the
resonance energy $\omega_{k}=\Omega$, leaving the TLS in the lowest
sublevel, and the other component has the resonance energy
$\omega_{k}=\Omega+\omega$, leaving the TLS in the second lowest
sublevel. As $\alpha$ increases, higher-order phonon processes take
effect. In consequence the scattered photon splits into a mixed
state, of which each pure state component is entangled with a
phononic state of the recoiled TLS and has a separate resonance
energy.

\begin{figure}
\begin{center}
  \includegraphics[bb=0 0 385 270, width=8cm]{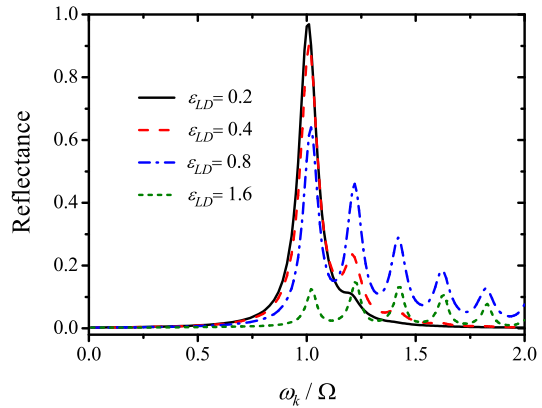}

\caption{\label{fig:compare_LD} (Color online) The reflection
spectra with $\varepsilon_{LD}=0.2$, $0.4$, $0.8$ and $1.6$ are
compared. The other two parameters are $\omega/\Omega=0.2$ and
$\Gamma/\Omega=0.05$.}
\end{center}
\end{figure}

To manifest the recoil effects, in Fig.\ref{fig:compare_LD} we
compare the reflection spectra as LDP varies. When
$\varepsilon_{LD}$ is very small, the reflection spectrum coincides
with the Lamb-Dicke limit. As $\varepsilon_{LD}$ increases, the
reflection peak at $\omega_{k}/\Omega=1$ is lowered and more peaks
emerge near the points $\omega_{k}=\Omega+n\omega$, with
$n=1,2,3\cdots$.

The emergent multi-peaks in the reflection spectrum can also be well
understood in an intuitive way. As shown in Fig.\ref{fig:level}, the
phononic excitations in the vibrational motion of the TLS bring
side-bands to the energy level structure of the motional TLS. As a
result, instead of the single transition between two levels for
fixed TLS, there are different transitions between the side-bands
and thus multi-peaks arise. For TLS initially in the lowest
sublevel, the transitions are shown by the dashed lines and the
corresponding resonance energies are $\Omega$, $\Omega+\omega$,
$\Omega+2\omega$. If the TLS is initially in a high sublevel, the
resonance energy may be $\Omega-\omega,\Omega-2\omega,\cdots$, as
shown by the dotted lines.

\begin{figure}
\begin{center}
  \includegraphics[bb=0 0 385 270, width=8cm]{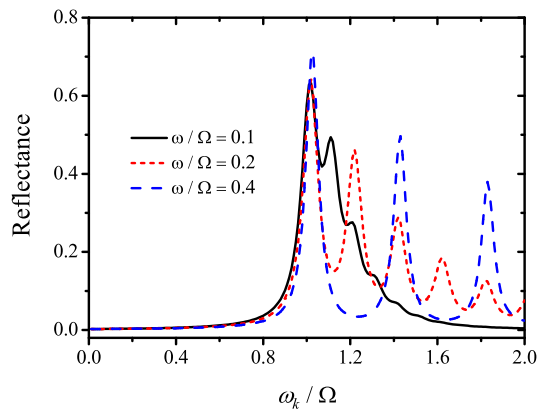}
\caption{\label{fig:compare_omega} (Color online) The reflection
spectra with $\omega/\Omega=0.1$, $0.2$ and $0.4$ are compared. The
other two parameters are $\Gamma/\Omega=0.05$ and
$\varepsilon_{LD}=0.8$. The reflection spectra are peaked near the
points $\omega_{k}=\Omega+n\omega$, with $n=0,1,2\cdots$.}
\end{center}
\end{figure}

The locations of the resonance peaks are illustrated in
Fig.\ref{fig:compare_omega}, which compares the reflection spectra
as $\omega/\Omega$ varies. Due to the interference between different
channels, which are distinguished by different phononic excitations
and correspond to a resonance energy respectively, the locations of
the reflection peaks have a small shift from the resonance energies
$\Omega+n\omega$, with $n=0,\pm1,\pm2,\cdots$. This phenomena can
also be interpreted as an energy level shift of the TLS resulting
from coupling between different sublevels via the photon field.

Besides multi-peaks in the reflection spectrum, which is a typical
effect due to the recoil of the TLS, the coupling strength is also
modified by the vibrational motion of the TLS. In the Lamb-Dicke
limit, the coupling strength is $J$, whereas for motional TLS, the
effective coupling is rescaled by the vibrational transition
amplitude, which varies with LDP, as written in
Eq.\eqref{eq:transition}. In the limit $\alpha\rightarrow\infty$,
the rescaling factor approaches zero, which means the coupling
between the photon and the TLS is negligible. As demonstrated in
Fig.\ref{fig:compare_LD}, the reflection diminishes with
$\varepsilon_{LD}$ increasing and in the limit
$\varepsilon_{LD}\rightarrow\infty$, the reflection vanishes.

\begin{figure}
\begin{center}
  \includegraphics[bb=0 0 385 270, width=8cm]{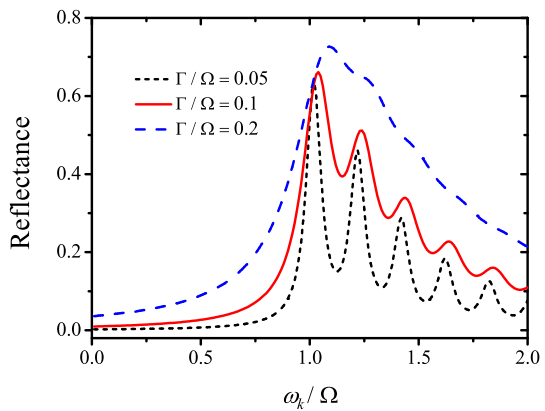}
\caption{\label{fig:compare_W} (Color online) The reflection spectra
with $\Gamma/\Omega=0.05$, $0.1$ and $0.2$ are compared. The other
two parameters are $\varepsilon_{LD}=0.8$ and $\omega/\Omega=0.2$.}
\end{center}
\end{figure}

The coupling strength between the photon field and the TLS is
indicated by the dimensionless parameter $\Gamma/\Omega$. As shown
in Fig.\ref{fig:compare_W}, when $\Gamma/\Omega$ increases, the
reflection increases and also, the width of the peaks increases. The
peaks merge into a single peak, when the width of the peaks is large
enough to fill out the spacing between them. In addition, the
location shift of the reflection peaks also increases with
$\Gamma/\Omega$ increasing.

Apart from the reflection spectrum, the recoil also shifts the
frequency of the outgoing photon. In the Lamb-Dicke limit, the
scattered photon is in a pure state described by the wavefunction

\begin{eqnarray}
u(x) & = & \theta(-x)\frac{e^{ik\cdot x}}{\sqrt{2\pi}}+\theta(-x)\frac{-i\Gamma}{\omega_{k}-\Omega+i\Gamma}\frac{e^{-ik\cdot x}}{\sqrt{2\pi}}\nonumber \\
 &  & +\theta(x)\frac{\omega_{k}-\Omega}{\omega_{k}-\Omega+i\Gamma}\frac{e^{ik\cdot x}}{\sqrt{2\pi}},
\end{eqnarray}
which shows that the frequency of the outgoing photon is still
$\omega_{k}$. In contrast, considering recoil effects, the scattered
photon is in a mixed state, of which the $n$th pure state component
is described by the wavefunction $\phi_{n}(x)$, whose long-distance
asymptotic behavior is
\begin{eqnarray}
\phi_{n}(x) & \sim & \theta(x)\frac{e^{ik_{n}\cdot x}}{\sqrt{2\pi}}t_{n}(k)\nonumber \\
 &  & +\theta(-x)\left(\delta_{n,0}\frac{e^{ik\cdot x}}{\sqrt{2\pi}}+\frac{e^{-ik_{n}\cdot x}}{\sqrt{2\pi}}r_{n}(k)\right),
\end{eqnarray}
with $k_{n}=k-n\omega/\upsilon_{g}$. It shows that the frequency of
the outgoing photon is $\omega_{k}-n\omega$, with $n\omega$ lost to
excite $n$ phonons in the vibrational motion of the TLS.

\section{Conclusions}

In this paper, we have studied the 1D single-photon scattering with
a motional TLS confined by a harmonic potential. In the Lamb-Dicke
limit when the trap frequency is very high or the mass of the TLS is
very large, we recover the results of previous studies
\cite{Fan,Zhou2008}, which assume the TLS is perfectly fixed and get
a reflection spectrum with only a single peak of complete reflection
at the resonance point $\omega_{k}=\Omega$. As the trap becomes
looser or the mass of the TLS becomes smaller, the recoil of the
motional TLS brings about new features to the single-photon
transport. The original single peak of complete reflection is
lowered and its location has a small shift from the resonance point
$\omega_{k}=\Omega$. At the same time, multi-peaks induced by phonon
excitations are observed near the points $\omega_{k}=\Omega+n\omega$
($n=0,\pm1,\pm2\cdots$) in the reflection spectrum. Besides, the
properties of the scattered photon are also changed. To be specific,
it splits into a mixed state due to its entanglement with the TLS's
motional state, and each component of the mixed state has a separate
frequency shift with respect to the incident photon, due to energy
exchange with phonon excitations.

In our forthcoming studies, this single-photon scattering model will
be generalized to some more realistic cases with higher spatial
dimension or multi-scatterers. For example, the three-dimensional
(3D) scattering of X-ray photons by a motional nucleon will be used
to explain the experimental results about coherent nuclear
scattering of synchrotron X-ray radiation.

\begin{acknowledgments}
This work is supported by National Natural Science Foundation of
China under Grants No.11121403, No.10935010, No.11222430,
No.11074305, No. 11074261 and National 973 program under Grants No.
2012CB922104.
\end{acknowledgments}

\end{document}